# Observation of near $E_F$ Fermi-arc van Hove singularity with prominent coupling to phonon in a van der Waals coupled Weyl semimetal

Z. Wang*[1,2], Cheng-Yi Huang*[3], Chia-Hsiu Hsu*[4,5,6], Hiromasa Namiki[7], Tay-Rong Chang[8], Feng-Chuan Chuang[5.6], H. Lin[3], Takao Sasagawa[7], Vidya Madhavan[2], and Yoshinori Okada[4]

[1]*Department of Physics, University of Science and Technology of China, Hefei, Anhui 230026, China.*
[2]*Department of Physics and Frederick Seitz Materials Research Laboratory, University of Illinois Urbana-Champaign, Urbana, IL 61801, USA*
[3]*Institute of Physics, Academia Sinica, Taipei, Taiwan*
[4]*Quantum Materials Science Unit, Okinawa Institute of Science and Technology (OIST), Okinawa 904-0495, Japan*
[5]*Department of Physics, National Sun Yat-sen University, Kaohsiung 80424, Taiwan*
[6]*Physics Division, the National Center for Theoretical Sciences, Hsinchu 30013, Taiwan*
[7]*Laboratory for Materials and Structures, Tokyo Institute of Technology, 4259 Nagatsuda, Midori-ku, Yokohama, Kanagawa 226-8503, Japan*
[8]*National Cheng Kung University, Taiwan*

A van der Waals (vdW) coupled Weyl semimetal material $NbIrTe_4$ is investigated by combining scanning tunneling microscopy/spectroscopy (STM/S) and first-principles calculations. We observe a sharp peak in the tunneling conductance near the zero bias energy, and its origin is ascribed to a van Hove singularity (vHs) associated with a Lifshitz transition of the topologically none trivial Fermi arc states. Furthermore, tunneling spectroscopy measurements show a surprisingly large signature of electron–boson coupling, which presumably represents anomalously enhanced electron–phonon coupling through the enhanced charge susceptibility. Our finding in vdW coupled material is particularly invaluable due to applicable exfoliation technology for searching exotic topological states by further modulating near $E_F$ vHs in nano-flake and its nano-device.

**Introduction**

The band structure near the Fermi energy ($E_F$) governs the physical properties of fermionic systems. A particularly interesting situation arises when there is a sharp peak in the density of states (DOS) near the $E_F$ [1,2,3,4,5,6,7,8,9,10,11,12,13,14]. One of the most exotic cases is that of high-temperature superconductor cuprates, in which the parent state is a half-filled two-dimensional (2D) square lattice [14]. Based on a simple tight binding without consider strong interactions, a van Hove singularity (vHs) exists near the $E_F$, and its important role of vHs for superconductivity has been suggested [14]. In general, whenever a band-structure-related sharp peak in the DOS coincides with the $E_F$, such as a vHs, it is intrinsically susceptible to Fermi surface instabilities, which can be a substantial driving force for various emergent phases, including unconventional superconductivity, density wave state, and magnetism, with the added possibility of multiple coexisting phases [14]. Dimensionality plays a critical role in Fermi surface instability because the band nesting condition and electron interaction are inherently enhanced by transitioning from three dimensions (3D) to 2D (and one dimension (1D)). Furthermore, the enhancement of electron–boson coupling, associated with enlarged susceptibility and Fermi surface instability, often becomes important [14,15]. Such phenomenology with band anomalies near the $E_F$ has been observed and extensively discussed for several material classes, such as van der Waals Dirac materials [5,6], Fe-based high-$T_c$ superconductors [7], magnetic Kagome systems [8,9,10], and strongly correlated oxides [11,12,13]. In particular, importance of van der Waals (vdW) coupled materials hosting near $E_F$ vHs has been growing due to applicable advanced exfoliation technology to further tailor vHs, as in graphene [5].

An interesting area in this field is the investigation of many-body effects in topological edge states. For example, prior experiments on topological crystalline insulators have revealed the vHs associated with a Lifshitz transition of the surface state [16]. However, the associated DOS peaks are far from the $E_F$ [17,18,19]. Other promising topological materials include the Weyl semimetals. Weyl semimetals feature pairs of doubly degenerate Weyl cones with a linear dispersion around the Weyl nodes in the bulk and Fermi arcs on the surface [20,21,22,23]. An important feature is that the surface Fermi arc states eventually morph into bulk-like states, that is, sinking into a pair of Weyl points with opposite charge chirality. Recent studies on the band structure of Weyl semimetals have suggested a possible topological Lifshitz transition near the Weyl points with an interesting surface Fermi arc that rewires at different energies [24,25,26]. However, a singularity with a peak in the DOS near the $E_F$ has not been directly investigated experimentally.

In this study, we combine scanning tunneling microscopy/spectroscopy (STM/S) and first-principles calculations to investigate a vdW coupled Weyl semimetal, NbIrTe$_4$. This ternary compound has attracted considerable interest as a type-II semimetal possessing relatively simple topological Fermi arc states with fewer Weyl points than those of MoTe$_2$ [27,28,29,30,31]. Using STM/S, we report the discovery of a near-$E_F$ vHs associated with a Lifshitz transition of the topological Fermi arc states in NbIrTe$_4$. Additionally, STS shows surprisingly prominent electron–boson coupling features, which could be a consequence of the sharp DOS peak from the topological vHs near the $E_F$.

## Results and Discussions

### Crystal structure and topographic image

The crystal structure of NbIrTe$_4$ from side and top views are shown in Figure 1a and b, respectively. The expected crystalline structure is 1T, and the termination is (001) surface with the van-der-Waals-coupled Te atomic sheet. A topographic image of NbIrTe$_4$ is shown in Figure 1c, and its inset shows zoomed-in image. Figure 1d shows the topographic height evolution along lines in the inset of Figure 1c. Compared with 1T-MoTe$_2$, it is expected that the lattice periodicity along the *b*-axis is larger in NbIrTe$_4$ owing to the alternate ordering of Nb and Ir rows (Figure 1b). The periodicity along *a* (green arrow) and *b* (purple arrow) directions are about 12.6 Å and 3.8 Å, respectively. Indeed, these values are consistent with the expected crystal structure of NbIrTe$_4$ (see arrows in Figure 1b and d).

### Conductance peak near zero bias voltage

The most remarkable experimental observation in this study is sharp dI/dV peak close to the zero bias voltage energy $eV_B$=0. Here, zero bias voltage corresponds to experimental E$_F$ position. To demonstrate this striking observation, we show a spatial evolution of line cut of the dI/dV along a line trace of approximately 350 nm (Figure 1c) and its spatially averaged spectra (Figure 1f). Importantly, the sharp dI/dV peak is homogeneously distributed on the surface. Thus, near zero bias energy dI/dV peak is not due to the disorder-originated local state. In addition, we confirmed that observation of near $eV_B$=0 dI/dV peak was reproducible by using different fresh tips, while the peak position was not always pined at $eV_B$=0 but its position differ within 10 meV scale depending on area by area separated by long distant (supplemental note 1 for details). As we introduce later, theoretical three energies for Weyl point (WP), to be compared with experiment, are indicated in Figure 1f. Also, near $eV_B$=0 spectral anomaly reflecting electron-boson coupling are indicated by arrows, as we discuss later.

### Experimental Quasi particle interference (QPI)

We performed quasiparticle interference (QPI) measurements to reveal the momentum origin of the experimental near E$_F$ DOS peak. QPI has been widely used to investigate the topological surface states in Weyl semimetals [32,33,34,35,36,37]. The dI/dV map at $eV_B$=+110 meV shows faint real-space scattering patterns confined sharply near the scattering centers, which represent quasiparticle states with short lifetimes (Figure 2a and its inset). On the other hand, the dI/dV map at $eV_B$ = +10 meV shows apparent wave-like patterns that can be tracked over several wavelengths (Figure 2b and its inset). Characteristic standing wave vectors can be visualized in the *q* space from the energy evolution of the Fourier transform (FT) of the dI/dV maps (Figure 2c-f). Figure 2g shows energy evolution of QPI channel along q$_x$, which is along *a*-axis in the real space and the Γ–Y direction in the momentum space (Figure 3a). As seen, dispersion is clearly observed in wide anergy range (form $eV_B$~0 up to 300 meV). An important feature to emphasize here is dramatic change of sharpness in QPI pattern between near $eV_B$=0 and high energy. While QPI channel is blunt at high energy (Figure 2c), near $eV_B$=0 QPI pattern shows multiple well-defined channels q1~q3 in FT image (Figure 2e). This trend in FT image is consistent with standing wave pattern in real space (inset of Figure 2 and Figure 2a-b). We note that NbIrTe$_4$ hosts much sharper interference pattern around $eV_B$=0, compared to the QPI observation in (Mo,W)Te$_2$ [33,34,35,36,37,38]. Hereafter, we present theoretical band structure in this system, followed by its comparison with experimental QPI dispersion.

**Density functional theory (DFT)**

To understand the band structure and topological nature of NbIrTe$_4$, we conducted density functional theory (DFT) calculation. Figure 3a shows Brillouin zone (BZ) of bulk (lower) and surface (upper). The three pairs of irreducible Weyl points (WP1, WP2, and WP3) are found in DFT calculation (see also supplemental note 2 for details). The pairs of Weyl points are reflected by mirror symmetry or glide symmetry (time-reversal symmetry) with opposite (equivalent) chiral charges. As shown in Figure 3b, irreducible WPs are depicted by color-coded circles and squares in the first BZ with their energies $E_{WP1}$, $E_{WP2}$, and $E_{WP3}$ (136 meV, 134 meV, and -86 meV, respectively). Figure 3c and d show the total-surface-spectral-weight maps at 134 and -50 meV, respectively. Here, it is known that the spectral weight of the topologically nontrivial Fermi arc surface state (NSS) sinks into the bulk state through Weyl points. Thus, by looking for this characteristic spectral fingerprint, NSS can be qualitatively disentangled with trivial surface state (TSS). The momentum region highlighted by the pink rectangle in Figure 3c clearly represents that spectra indicated by arrow (NSS2) sinks into the bulk bands through WP2, as shown in Figure 3e. Similarly, tracking spectral sinking into WP3 enable us to assign another Fermi arc portion. Figure 3f shows dispersion along the vertical line with an arrow in Figure 3d, and spectral feature indicated by arrow (NSS3) sinks into the bulk bands through WP3. Here, the spectral weight sinking behavior into WP2 and WP3 in the momentum space are highlighted by the dashed curve in Figure 3e and f, respectively.

We briefly point theoretical observation of unusual connection of NSS2 and NSS3 in momentum space and their characteristic atomic nature. Figure 3g shows energy evolution of total surface spectral weigh (NSS and TSS) along $k_x$ at fixed $k_y=\pi/b$ (see horizontal arrows in Figure 3c). Around the momentum highlighted by dashed circle, NSS2 and NSS3 merge with each other. This means that NSS2 and NSS3 host a single Fermi arc state, which connect two pairs of Weyl points (a total of four Weyl points) instead of the conventional case where only one pair of Weyl points is connected. This behavior is topologically allowed since number of nontrivial surface states emerging from each Weyl node with chiral charge-1 remains one. While we leave understanding details for future theoretical study, we think it is possible that this connection is not accidental. Another, feature we find from band calculation is dominant contribution of Nb atom for constructing NSS (see also supplemental note 3 for details). Figure 3h is atomic character dependence of spectral weight dispersion shown in Figure 3g. Dominant contribution of Nb atom for NSS2 and NSS3 are clearly visualized (see arrows). In this report, we simply identify the NSS from TSS, based on the relative spectral weigh of Nb than that of Ir/Te. The ambiguity of threshold does not have an issue within the main purpose of this study.

Calculated surface spectral weight shows existence of a Lifshitz transition of nontrivial Fermi arcs at characteristic energy. We define this characteristic energy as $E_{vHs}$, whose theoretical value is around -60 meV in pristine DFT calculation (see vertical axis of Figure 3g and h). By focusing on the momentum region highlighted by rectangle in Figure 3d, the detailed energy evolution of the Fermi arcs within a small energy interval near the $E_{vHs}$ are shown in Figure 4a-e. The interaction and eventual merging of the two bands result in a saddle point in the dispersion, that is, the formation of a vHs. As highlighted by the dashed circle in Figure 4c, we observe a geometrical change in the constant energy contour shape at $E_{vHs}$, which is the Lifshitz transition happening at this energy. The calculated DOS of the NSS exhibits an apparent peak at the corresponding energy $E_{vHs}$ (Figure 4f). To further check momentum origin of this peak, we plot the calculated dispersion of the NSS (upper panel in Figure 4g) by zooming in around the momentum of band touching, together with a map of $1/|\nabla_k E|$ in the momentum space (lower panel in Figure 4g). As the total DOS is proportional to $\int 1/|\nabla_k E(k)| dk$, the $1/|\nabla_k E|$ map provides an excellent way to visualize the characteristic momentum leading to an enhanced DOS. As shown in Figure 3g, an apparent correspondence can be observed between the Lifshitz transition (upper panel) and the large DOS (lower panel). A next important step is matching energy between theory and experiment to address the momentum origin of the experimental DOS peak near $E_F$. Adjusting energy between theory and experiment is ordinary required process due to miss estimation of $E_F$ in DFT calculation and/or inherent chemical doping/displacement in $NbIrTe_4$ phase. While clarifying microscopic origin is beyond the scope of this study, natural expectation for possible chemical origin is Te vacancy and/or displacement between Ir and Nb.

**Comparison between experiment and theory**

Based on agreement between experimental QPI and theoretical joint density of state (JDOS) patterns, we conclude that experimental near $E_F$ DOS peak originates from vHs of NSS. We first show comparison between experimental QPI (Figure 5a) and JDOS from both NSS and TSS (Figure 5b). Note that energy axis of theoretical JDOS is set as $E-E_{vhs}$. Clear agreement seen in this comparison strongly supports that experimental $eV_B = 0$ roughly corresponds to theoretical $E_{vhs}$. Furthermore, theoretical spectral weight distribution of TSS and NSS near $E_{vHs}$ and experimental $E_F$ are beautifully captured by experimental QPI pattern. To demonstrate this, theoretical spectral weight distribution in $k_x$-$k_y$ plane for $E-E_{vHs}=10$ meV, as well as scattering vector candidates ($q_{tss}$, $q_{nss1}$, $q_{nss2}$, and $q_{nss3}$) is shown first (Figure 5c). Focusing on 10 meV other than $E_F$ is simply due to higher experimental signal to noise, which brings much cleared comparison, and the momentum area highlighted by rectangle in Figure 5c is where we focus in Figure 4 to demonstrate vHs associated with Lifshitz transition. By comparing QPI patterns at $eV_B = 10$ meV (Figure 5d upper) and theoretical JDOS sorely from NSS at $E-E_{vHs}=10$ meV (Figure 5d lower), experimental three characteristic channels $q_1$, $q_2$, $q_3$, and their extension in $q_x$-$q_y$ space are beautifully explained as intra-scattering channel $q_{nss1}$, $q_{nss2}$, and $q_{nss3}$, respectively. Since JDOS simply captures QPI, we expect spin-based matric element is not totally diagonal for $q_{nss1}$, $q_{nss2}$, and $q_{nss3}$. Furthermore, absence of $q_{tss}$ means our QPI data are dominated by the NSS near the $E_F$. Note that TSS contains either Te or Ir components while NSS is dominated Nb component. Thus, for understanding missing experimental $q_{tss}$, we think the orbital related matrix element in scattering processing play a dominant role instead of the selection of spin-based matrix element effect. Based on comparison between theory and experiment mentioned here, we conclude that experimental dI/dV originates from vHs associated with the Lifshitz transition of NSS.

**Experimental electron-boson coupling**

Interestingly, prominent spectroscopic signature of electron boson coupling appears in experimental dI/dV near $E_F$. Existence of prominent electron boson coupling is along with the fact that slight quantitative mismatch between experimental scattering vectors and theoretical JDOS without considering interaction (see Figure 5). Figure 6a shows a topographic image in which 2116 tunneling spectra were acquired near the $E_F$. The averaged spectra are shown in (Figure 6b). Peak-dip features, which are symmetric with respect to the $E_F$, are observed in the tunneling spectrum (dI/dV) and its derivative ($d^2I/dV^2$), indicating a strong coupling of electrons to a bosonic mode. Here, we define $E_p$ as the DOS peak energy and $E^+$ and $E^-$ as the peak positions in $d^2I/dV^2$. The area in Figure 6a and Figure 1-2 are different, and this presumably lead slight difference of $E_p$ in Figure 1f and Figure 6b (see also supplemental note 1 for details).

We observe periodic modulations in the intensity of the DOS peak and features related to the electron–mode coupling. To present this variation, we simply show the spatial evolution of the topographic height (Figure 6d), dI/dV (Figure 6e), and $d^2I/dV^2$ (Figure 6f) across lines A-B in Figure 5a. It can be observed that both the peak intensity at the $E_P$ and the features at $E^+$ and $E^-$ vary in real space with the crystal lattice. These spatial variations of peak intensity at $E_p$ with lattice periodicity can be naturally understood, as our DFT calculation indicates that the dominant contribution in the formation of the vHs of the NSS is from Nb (Figure 3h and supplementary note 3). Importantly, $E^+$ and $E^-$ appear almost symmetrically relative to the $E_F$ at every location, which further supports that these features at $E^+$ and $E^-$ are related to the electron–mode coupling. As our system is not magnetic, the bosonic mode is more likely to be a phonon. Indeed, the energy scale of $|E^+|$ and $|E^-|$ is comparable to that of the calculated phonon energy while our current comparison is with bulk phonon, leaving direct comparison with surface phonon for future study (supplemental note 5 for details). Note that we observed no evidence of gap opening and an additional order on the surface of $NbIrTe_4$ despite the presence of a significant peak in the DOS near the $E_F$. This is rather intriguing, given that strong electron–boson coupling can be revealed from the STM data.

A spectroscopic observation of prominent electron–phonon coupling with a topological vHs is not trivial. For example, in the case of $Z_2$ topological insulators, electron–phonon coupling is not observed to have a large signal [39,40,41,42]. In contrast, a recent transport study of a Weyl semimetal, $WP_2$, has reported dominant electron–phonon interactions [43]. While the nature of the surface has not been clarified, our observations suggest an intriguing interplay between the topological vHs formation and enhanced

electron–phonon coupling. Notably, our experimentally observed DOS peak is much sharper than that obtained from a noninteracting model (Figure 4f), which encourages microscopic theoretical studies for a detailed understanding of the many-body effects in topological Fermi arc states in Weyl semimetals. Furthermore, our findings provide an interesting route to modulate the Fermi arc states via phonon engineering [44,45,46,47,48]. While phono is reasonable origin, as discussed possible other bosonic mode coupling is not totally excluded. For example, acoustic plasmon mode is reported recently in $Bi_2Te_3$ [49], and possible multiple nature of bosonic mode coupling may further enrich many-body effects on near $E_F$ topological vHs.

**Summary**

In summary, we report the discovery of formation of a DOS peak near the $E_F$ on the surface of the Weyl semimetal $NbIrTe_4$. Combined with the experimental observation of sharp QPI patterns and theoretical calculations, we find the origin of the experimental near-$E_F$ DOS peak is the formation of a vHs associated with the Lifshitz transition in topologically none-trivial Fermi arc states. Our observation indicates the intriguing interplay between the formation of the near-$E_F$ topological vHs and enhanced electron–phonon coupling, which paves the way for the manipulation of topologically nontrivial states in Weyl semimetals via phonon engineering. Our finding in vdW coupled material is particularly invaluable due to applicable exfoliation technology for searching exotic topological states by further modulating near $E_F$ vHs in nano-flake and its nano-device.

## Methods

### Crystal growth and STM experiments

The single crystals NbIrTe$_4$ were grown by a Te-flux method. The samples were characterized by XRD and XRF to confirm single phase without phase segregation. Samples were cleaved in ultra-high vacuum at about 80 K and then immediately inserted into the cold STM head. All the data were obtained around 300 mK in zero external magnetic field. Different W and PtIr tips were used and checked on Cu(111) surface. The tunneling conductance (dI/dV), which is proportional to DOS, were measured using standard lock-in technique with voltage modulation at 987.5 Hz.

### The density functional theory (DFT) band calculation

The density functional theory (DFT) band calculation was performed using the projector augmented wave method as implemented in the VASP package [50] within the generalized gradient approximation (GGA) [51]. In all of calculation, the spin–orbit coupling (SOC) was included self-consistently.

### Acknowledgements

Work at the University of Illinois, Urbana-Champaign was supported by U.S. Department of Energy (DOE), Office of Science, Office of Basic Energy Sciences (BES), Materials Sciences and Engineering Division under Award # DE-SC0022101. Z.Y.W. is supported by National Natural Science Foundation of China (No. 12074364) and the Fundamental Research Funds for the Central Universities (WK3510000012). This work was, in part, supported by a JST-CREST project [JPMJCR16F2] and a JSPS Grants-in-Aid for Scientific Research (A) [21H04652].

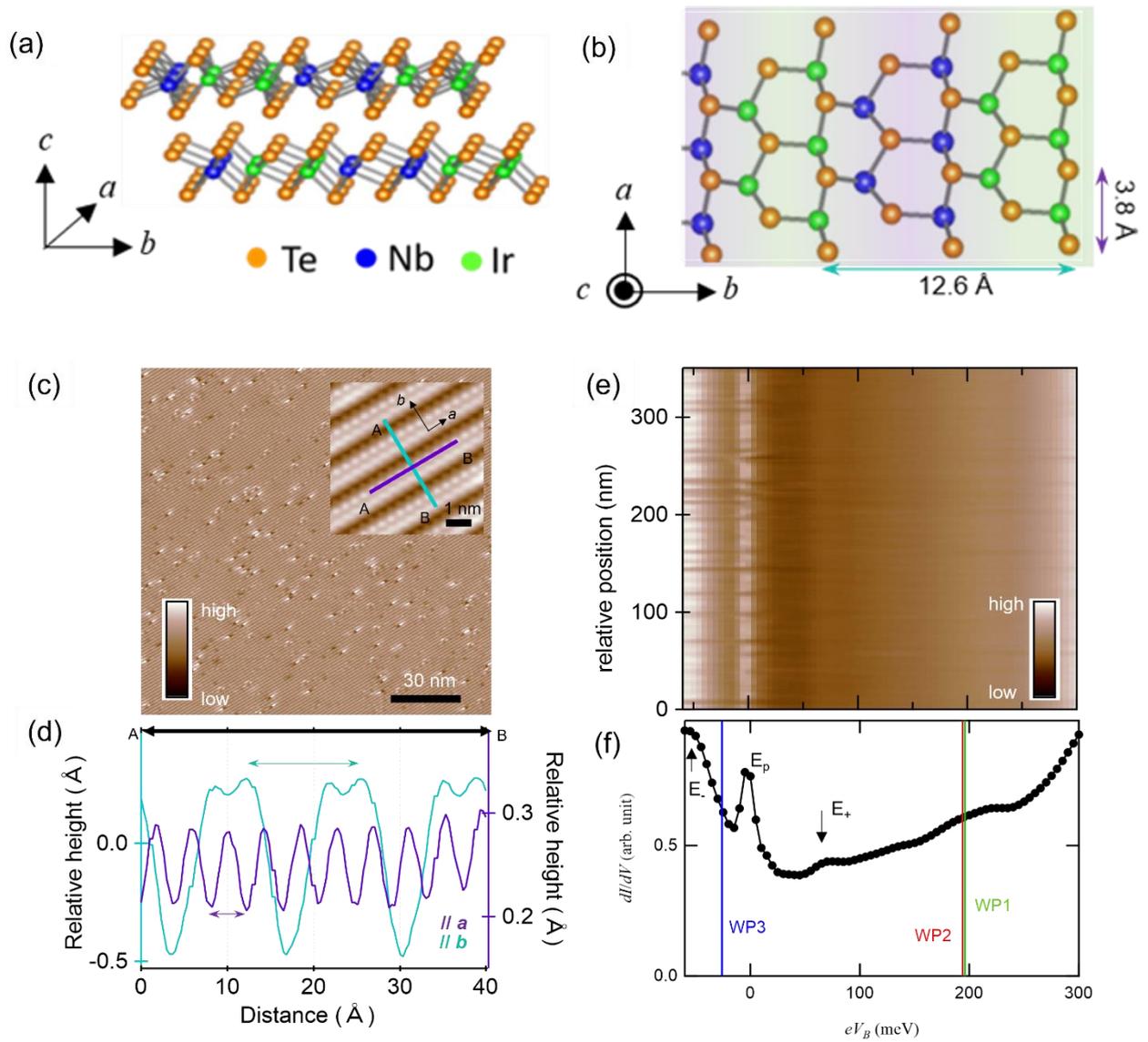

**Figure 1. Characteristics of the Weyl semimetal NbIrTe$_4$.**
**(a-b)** Schematic of the crystal structure of NbIrTe$_4$ with side and top views, respectively. **(c)** STM topographic image ($I_t$=100 pA and $V_B$=100 mV). **(d)** Topographic height variations along two lines in the inset of (c). **(e)** Homogeneous spatial evolution of dI/dV spectra along a 350 nm line on the surface, and **(f)** the spatial average of dI/dV curve.

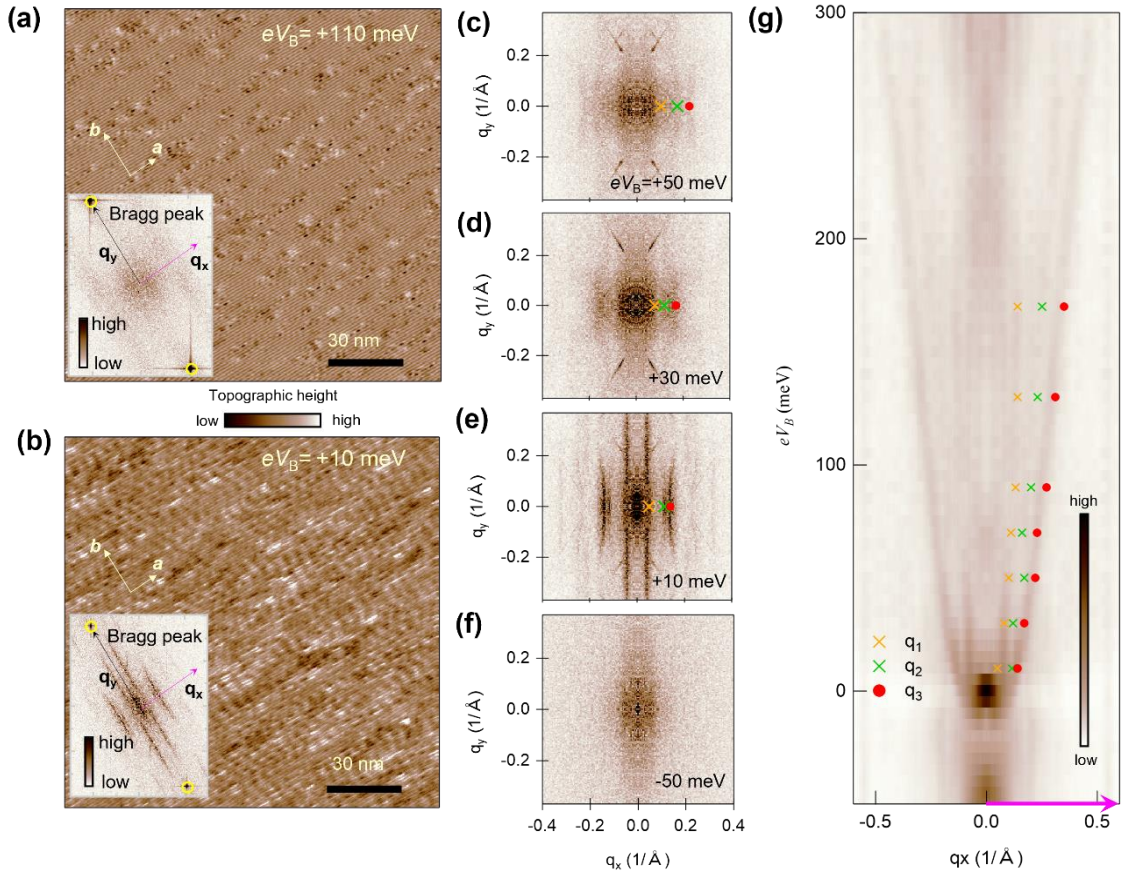

**Figure 2. QPI patterns of NbIrTe$_4$ surface.**

**(a-b)** Conductance **(**d*I*/d*V***)** maps at $eV_B$= 110 and 10 meV, respectively. The insets show the raw Fourier transformation (FT) of the d*I*/d*V* images, where Bragg peaks are highlighted by circles. **(c-f)** Energy evolution of FT of d*I*/d*V* maps, which are zoomed in around q=0 to emphasize the interference patterns. Here, the horizontal and vertical axes are set along the crystal axes *a* and *b*, respectively. Also, these images are rotated and two-fold symmetrized. **(g)** Intensity plot of the line cut in the FTs along the q$_x$ direction at q$_y$=0, together with scattering vectors.

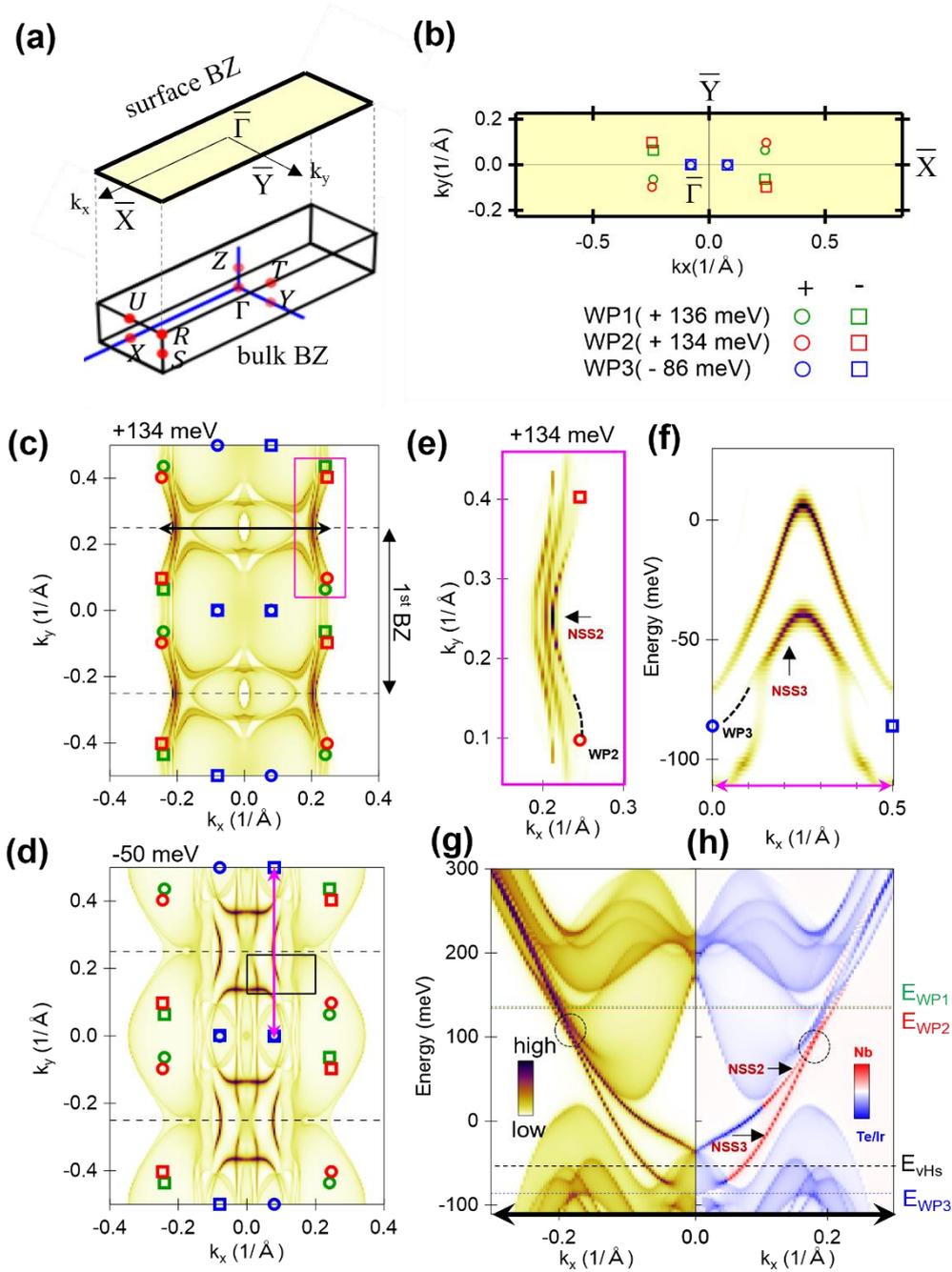

**Figure 3. The Density functional theory (DFT) calculation for electronic state of NbIrTe4.**

**(a)** BZ of the system. **(b)** The calculated WPs with their momentum, energy, and chirality. **(c-d)** Total spectral weight on the surface plotted at 134 meV, the energy of WP2, and near the Fermi energy at -50 meV. The total surface spectral weight contains TSS and NSS contributions. The Weyl points WP1 (green), WP2 (red), and WP3 (blue) are indicated in the momentum space. **(e)** Zoomed-in image of the pink rectangle in (c), and **(f)** dispersion along the pink line in (d). The dashed lines indicate the trace along which the NSSs sink into the bulk state through Weyl points in (e) and (f). **(g)** Energy dispersion along the $k_x$ direction at $k_y=\pi/b$ [see horizontal line in (c)] for total spectral weight and **(h)** its atomic character dependence.

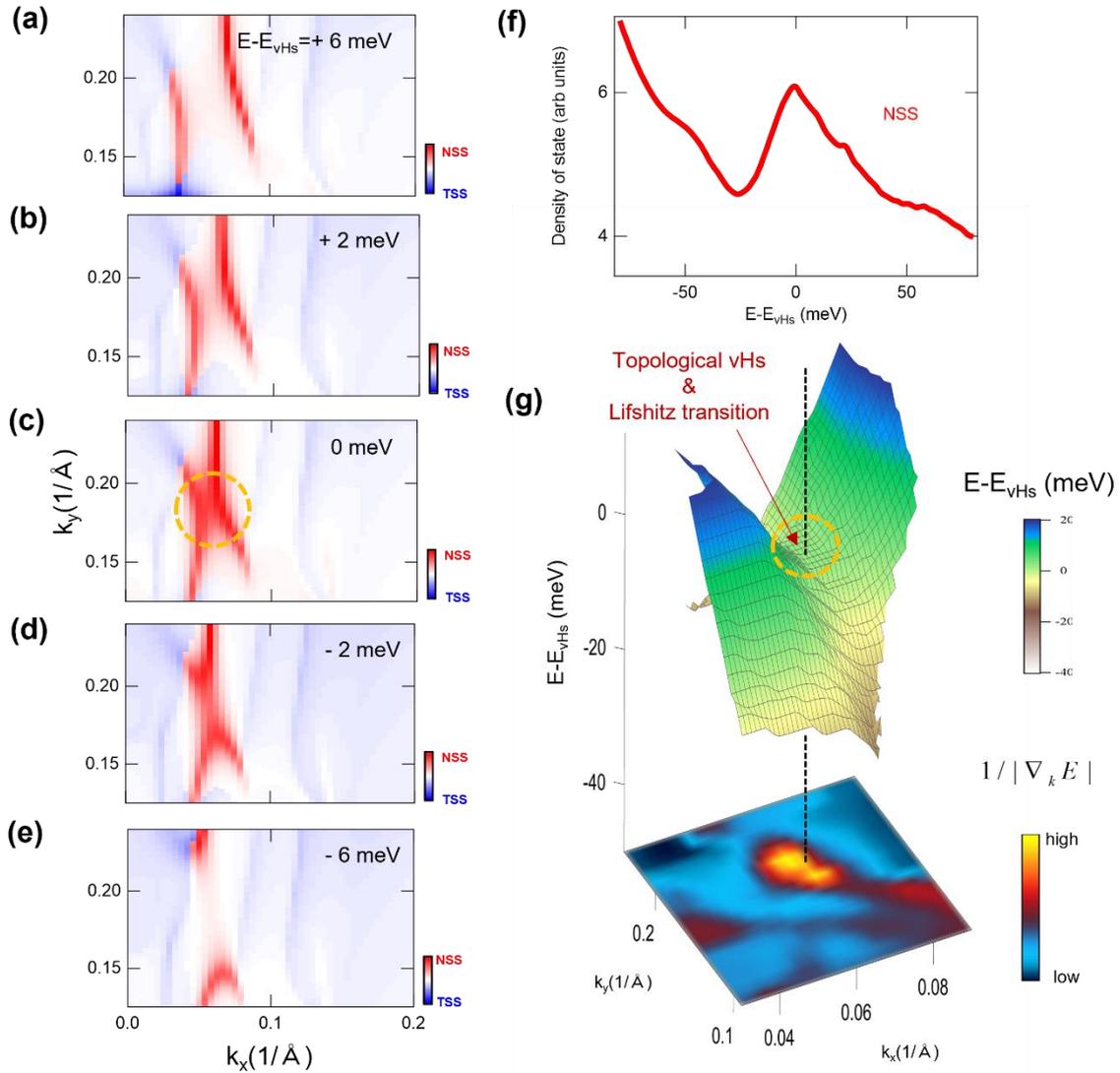

**Figure 4. Electronic structure calculation near the energy for van hove singularity $E_{vhs}$.**
**(a-e)** Energy evolution of spectral weight for the area indicated by the dashed rectangle in **Fig. 3(d)**. **(f)** Calculated DOS for NSS, which is obtained by integrating the spectral weight over all the momenta in the first BZ. **(g)** 3D display of NSS dispersion (upper part) and $1/|\nabla_k E|$ (lower part), near the critical momentum (dashed circle in (c)). The vHs formations associated with a Lifshitz transition are indicated by an arrow and dashed circle in (c) and (g).

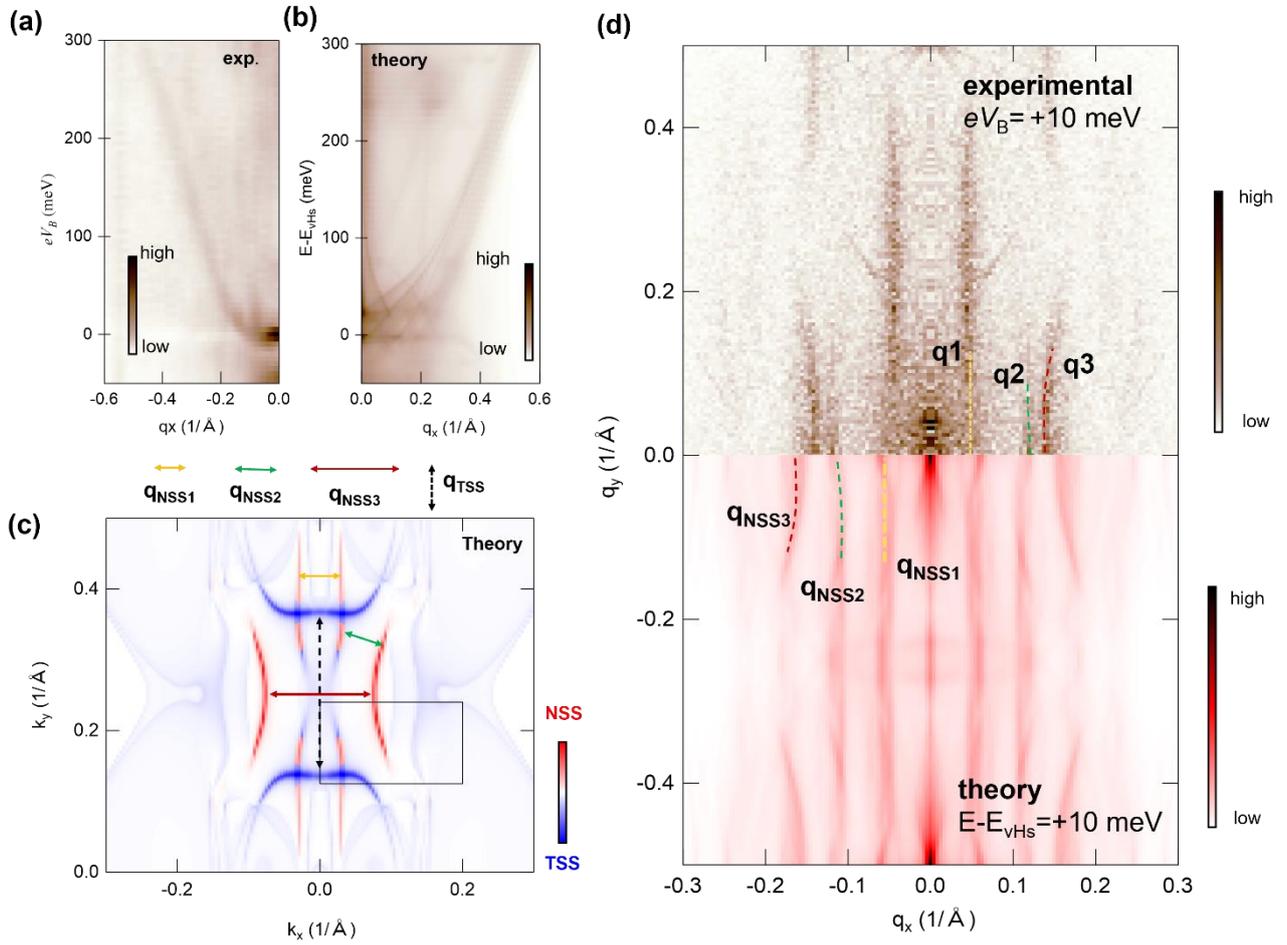

**Figure 5. The comparison between theory and experiments**

(a) The experimental energy evolution of QPI patter along qx direction. (b) The theoretical autocorrelation of the total spectral weight (JDOS) along the $q_x$ direction at $q_y$=0. Here, vertical axis is plotted energy relative to vHs, which is $E-E_{vHS}$. (c) Calculated spectral weight at $E-E_{vHs}$ =10 meV. Here, the rectangle is the momentum area focused in detail in Fig. 4. (d) Comparison between the experimental interference pattern (upper) and calculation based on NSS (lower) at $eV_B$= 10 meV and $E-E_{vhs}$=10 meV, respectively.

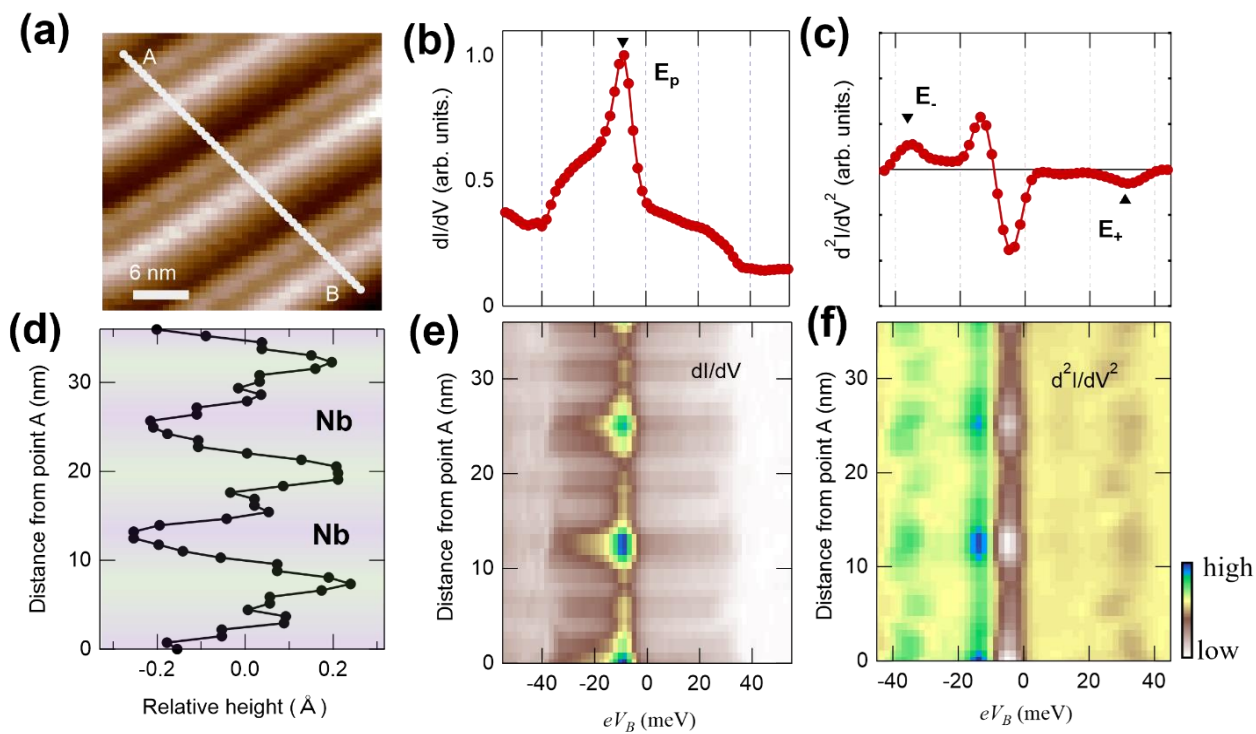

**Figure 6. Enhanced electron–mode coupling.**

**(a)** Topographic image ($I_t$=100 pA and $V_B$=100 mV). **(b-c)** Spatial average of dI/dV and $d^2I/dV^2$ spectra. The spectra are averaged from 2116 curves (46 x 46 pixels) from the area shown in (a). **(d)** Topographic height variations along A-B in (a). **(e-f)** Spatial evolution of dI/dV and $d^2I/dV^2$ spectra taken along A-B shown in (a).